\documentstyle[psfig]{mn}

\newif\ifAMStwofonts

\def\lapp{\ifmmode\stackrel{<}{_{\sim}}\else$\stackrel{<}{_{\sim}}$\fi}
\def\gapp{\ifmmode\stackrel{>}{_{\sim}}\else$\stackrel{>}{_{\sim}}$\fi}
\def\psr{PSR~B0540$-$69}
\def\arcdeg{\hbox{$^\circ$}}

\title[Observations of PSR B0540--69]
{Radio and X-ray Observations of PSR B0540--69}
\author[Johnston et al.]
{Simon~Johnston$^1$, Roger~W.~Romani$^2$, F.~E.~Marshall$^3$ \& W.~Zhang$^3$\\
$^1$School of Physics, University of Sydney, NSW 2006, Australia.\\
$^2$Dept. of Physics, Stanford University, Stanford,
CA  94305-4060, USA.\\
$^3$Laboratory for High Energy Astrophysics, Goddard Space Flight Center,
Greenbelt, MD 20771, USA.\\
}
\date{\today}
\pagerange{\pageref{firstpage}--\pageref{lastpage}}
\pubyear{2004}
\begin{document}
\maketitle
\label{firstpage}

\begin{abstract}
\psr{} is one of a small handful of pulsars known to emit giant pulses
and the only extra-galactic pulsar known to do so. We observed the
pulsar for a total of 72~h over a 6 month interval and detected
141 giant pulses. We have obtained correct phasing between the radio
arrival times of the giant pulses and the X-ray pulse profile.
The giant pulses occur in two phase windows, located 6.7~ms before
and 5.0~ms after the midpoint of the X-ray profile. We have detected
the integrated profile of the pulsar at 1.4~GHz with a flux density
of only 24~$\mu$Jy. The statistics of the giant pulses are clearly power-law and
it is likely that the giant pulses contribute only a few per cent to the
integrated pulse flux density, similar to other pulsars with giant pulse
emission.
Simultaneous X-ray and radio observations show
no significant increase in the X-ray flux at the time of the radio giants.
The relative enhancements in the X-ray emission must be at least 5.5 times
smaller than in the radio.
\end{abstract}


\section{Introduction}
\psr{} was discovered in the early 1980s by Seward, Harnden \& Helfand
(1984) using data from the Einstein X-ray Observatory.
The pulsar is located inside the supernova remnant SNR 0540--693
in the Large Magellanic Cloud. It has a short rotation period ($\sim$50~ms)
and a rapid spindown with a characteristic age of only 1500~yr.
The X-ray profile of the pulsar consists of a single broad profile
which covers about half the pulse phase \cite{pkh03}. The profile
is not Gaussian in shape but appears to be double peaked and also
contains structure on the rising and trailing edges. The profile
does not appear to evolve significantly from optical to hard X-rays.
This pulse shape is in contrast to the high-energy emission from
the Crab pulsar. In the Crab, two sharp peaks with a separation
of 0.4 pulse phase are seen at all energies.

The radio emission from \psr{} went undetected for a decade until
Manchester et al. (1993)\nocite{mml+93} discovered a
broad, weak radio pulse at 0.64 GHz. The flux density
at that frequency is 0.4~mJy and the pulse duty cycle is 
more than 80\% with the hint of a double profile, similar to the
profile seen at high energies.
Manchester et al. (1993)\nocite{mml+93}
failed to detect the pulsar at either 1.4 or 0.44 GHz.

The pulsar was observed with the Parkes radio telescope in 2001
as part of a survey to detect giant pulses from
young and millisecond pulsars. Single pulses were detected 
with energy more than 1000 times that of the
average pulse energy (Johnston \& Romani 2003\nocite{jr03}; hereafter Paper~I).
Such strong pulses, first seen in the Crab pulsar, are called `giant pulses' 
and this was their first detection from an extra-galactic pulsar.
The giant pulses are
scatter broadened at 1.4 GHz with an exponential scattering time
of 0.4~ms and have an emission bandwidth of at least 256 MHz.
There is some evidence that the flux density distribution of the giant
pulses is a power-law with a shallower index than seen for the
Crab and PSR~B1937+21 (Paper~I).
In the Crab pulsar, PSR~B1937+21 and PSR~B1821--24 the giant pulses 
are exactly in phase with the high energy emission which provides evidence
that their emission may have a common origin \cite{rj01}.
An attempt was made to match the phases of the radio giant pulses 
with the X-ray profile for \psr{} obtained with the
Rossi X-ray Timing Explorer (RXTE), however this was subject 
to caveats concerning
the absolute timing between the X-ray and the radio, and, as is shown later 
in this paper, was incorrect in Paper~I.
In 8~hr of integration Johnston \& Romani (2003) failed to 
detect any integrated flux density from the pulsar to 
a level of 13~$\mu$Jy, assuming a duty cycle of 10\%. This implies the spectral
index between 0.64 and 1.38~GHz must be steeper than --4.4.

Timing of the pulsar has been carried out in the X-ray band since its
discovery. The two most recent papers dealing with the timing are
Cusumano, Massaro \& Mineo (2003)\nocite{cmm03}
and Zhang et al. (2001)\nocite{zmg+01}. In the latter paper,
the authors found evidence for a glitch in the pulsar, however,
Cusumano et al. (2003) ascribe this to timing noise. The main result
of the timing of this pulsar is that 
the pulsar braking index is substantially less than the
value of 3.0 expected from pure magnetic dipole spindown.
The actual value of the braking index is subject to some uncertainty
and ranges from 1.8 to 2.8 depending on the data set used.

Simultaneous searches for giant pulse emission at radio and high energies
have been made towards the Crab pulsar. Lundgren et al. (1995)\nocite{lcu+95}
observed with OSSE in the 50 - 220 keV band and in the radio. They detected
no enhanced emission at the high energies and concluded that the
$\gamma$-ray enhancement is a factor of 10 less than in the radio.
More recently, Shearer et al. (2003)\nocite{sso+03} reported a small
increase in the optical flux of the Crab coincident with the giant
pulses in the radio. However, the optical enhancement is 700 times
less than seen in the radio.
Both Vivekanand (2001a,b)\nocite{viv01a,viv01b} and
Cusumano et al. (2003)\nocite{chk+03} report on long RXTE observations
of the Crab and PSR~B1937+21 respectively and reported limits on
X-ray enhancements at least 640 and 230 times less than expected
from equivalent duration radio observations.

In this paper we report on further radio observations of \psr{}.
We obtained simultaneous X-ray and radio observations on 2003 August 6
and September 28.
In conjunction with data from the Crab pulsar, this has allowed us
to directly compare the phase of the X-ray profile with the radio giants.
This is described in detail in Section 3. In Section 4 we discuss
the location and statistics of the giant radio pulses,
and in Section 5 we report on the detection of the integrated
profile of the pulsar. We conclude with some comments on the implications
for the location and mechanism of the magnetospheric process generating the
giant pulse emission.

\section{Observations}
All the radio observations of \psr{} were made with the 64-m Parkes
radio telescope. We used the centre beam of the 21-cm multi-beam receiver
at an observing frequency of 1390~MHz. The receiver had a system equivalent
flux density of 27~Jy on cold sky. The back-end consisted of a filterbank
system containing 512 channels per polarization, each of width 0.5 MHz for
a total bandwidth of 256~MHz.  The polarization pairs were summed to
form the total power, each
output was then sampled at 80 $\mu$s, one-bit digitized,
and written to DLT for off-line analysis.
Table~\ref{obstab} lists the observation dates for the pulsar,
including the initial observations in 2001. 
The final column of the table gives the time on source during
the observing session. The total observing time for the 2003 
sessions was 71.5 h.
During the 2003 August and September observing
sessions the Crab pulsar was observed on each day for 20 min. The integration
time is sufficient to detect the integrated profile of the Crab and
a number of giant pulses.  This provides
us a check both for the software detection algorithm used in detecting
giant pulses and for the absolute timing between the
Parkes telescope and the RXTE satellite.
\begin{table}
\caption{Log for the radio observations of \psr{}.}
\begin{tabular}{llr}
\hline & \vspace{-3mm} \\
\multicolumn{1}{c}{Date Range} &
\multicolumn{1}{c}{MJD Range} & \multicolumn{1}{c}{Time (h)} \\
\hline & \vspace{-3mm} \\
2001 May 20 $-$ 22       & 52049 $-$ 52051 & 8.5 \\
2003 May 17 $-$ 18       & 52776 $-$ 52777 & 6.8 \\
2003 May 24 $-$ 25       & 52783 $-$ 52784 & 9.7 \\
2003 June 3              & 52793           & 4.3 \\
2003 August 4 $-$ 10     & 52855 $-$ 52861 & 31.2 \\
2003 September 7 $-$ 8   & 52889 $-$ 52890 & 10.3 \\
2003 September 24 $-$ 28 & 52906 $-$ 52910 & 9.2 \\
\hline & \vspace{-3mm} \\
\end{tabular}
\label{obstab}
\end{table}

In the off-line analysis the data were de-dispersed using
a dispersion measure (DM) of 146.5~cm$^{-3}$pc (Paper~I) to produce
an output stream of 80~$\mu$s time samples. Data showing strong
interference were clipped, and the data set searched for giant pulses
exceeding some threshold using the technique described in Paper~I.
For this part of the analysis, the absolute phase of each time sample
is not important.

In order to produce the integrated pulse profile, and to determine
the phase of the giant pulses, an accurate radio
ephemeris was first derived using standard timing techniques.
Then, each time sample could be tagged with
an appropriate pulse rotation phase. These samples were summed to form
an integrated profile with a maximum of 630 phase bins per period.

X-ray observations are obtained using RXTE.
\psr{} is observed regularly ($\sim$40 times per year) along with
the other young LMC pulsar J0537$-$6910 as part of a monitoring program
(Zhang et al 2001), as is the Crab pulsar.
X-ray arrival times are obtained from 3-20~keV PCA data,
using pulse height channels 5-50 from the top layer of the detector. 
After fitting to a sinusoidal template, these data provide arrival
times with an accuracy of $\sim$0.4~ms relative to UTC.
We used observations of the Crab taken near the epoch of the \psr{}
observations to cross-check the Parkes-RXTE timing (see section 3.1 below).

\begin{figure}
\centerline{\psfig{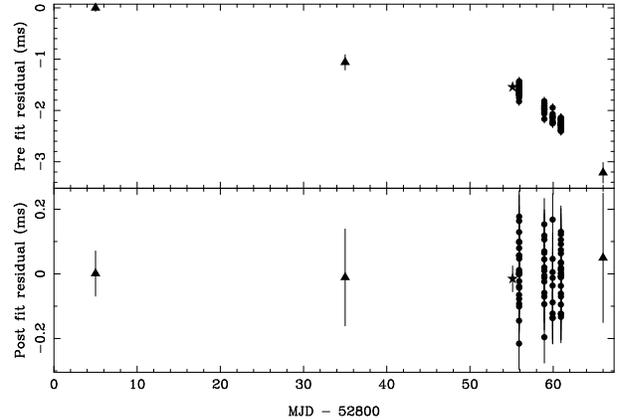}}
\caption{Pre fit (top panel) and post fit (bottom panel) residuals for the
Crab pulsar from radio data taken at Jodrell Bank and Parkes and X-ray
data from RXTE. The RXTE data point is marked with a star, the Jodrell Bank
data are shown as triangles and the Parkes data are shown as dots.}
\label{crab_fit}
\end{figure}

\section{X-ray and radio timing}
\begin{table*}
\caption{Timing solutions for \psr{}.}
\begin{tabular}{llll}
\hline & \vspace{-3mm} \\
                             &\multicolumn{1}{c}{Late giants}
                             &\multicolumn{1}{c}{Early giants}
                             &\multicolumn{1}{c}{X-ray}   \\
\hline & \vspace{-3mm} \\
R.A. (J2000)                 &  05$^{h}$40$^{m}$11$^{s}$.22
                             &  05$^{h}$40$^{m}$11$^{s}$.22
                             &  05$^{h}$40$^{m}$11$^{s}$.22\\
Dec. (J2000)                 &  $-$69$\arcdeg$19$\arcmin$54$\arcsec$.98
                             &  $-$69$\arcdeg$19$\arcmin$54$\arcsec$.98
                             &  $-$69$\arcdeg$19$\arcmin$54$\arcsec$.98\\
DM (cm$^{-3}$pc)             &  146.5 &  146.5 & -- \\
$\nu$ (s$^{-1}$)             &  19.775529613(6) & 19.775529618(6) & 19.7755296113(8)\\
$\dot \nu$ $(\times 10^{-12}$ s$^{-2})$ & $-$187.287(1) & $-$187.288(1) & $-$187.2853(4)\\
$\ddot\nu$ $(\times 10^{-21}$ s$^{-3})$ & 5.0(1.5) & 4.3(1.3) & 4.18(6)\\
Braking index, n             & 2.8(0.8) & 2.4(0.7) & 2.36(3)\\
Period epoch (MJD)           &  52857.866 & 52857.866 & 52857.866\\
No. TOAs                     & 43 & 32 & 42\\
Rms residual (ms)            & 0.91 & 0.98 & 0.48\\
Arrival phase                & 0.10 & 0.87 & 0.00\\
Offset from X-ray (ms)       & $+$5.0(0.3) & $-$6.7(0.3) & 0.00\\
\hline & \vspace{-3mm} \\
\end{tabular}
\label{timtab}
\end{table*}
\subsection{Crab}
In order to compare the absolute phasing between the RXTE X-ray 
observations and the Parkes radio observations, we first used data taken
on the Crab pulsar. The Crab was observed with RXTE on 2003 August 4
and the arrival time of the main pulse at infinite frequency at
the barycentre occurred 10938.05646366 seconds after MJD 52855.0.
At Parkes, observations of the Crab were made on 2003 August 4, 7, 8 and 9.
Each observation was 20~min in duration at a frequency of 1390~MHz
and the sampling rate was 80~$\mu$s. Topocentric arrival times at 1390~MHz
were recorded for the $\sim$20 largest giant pulses 
in each of the observations.
We also made use of the Jodrell Bank timing data on the Crab pulsar,
in particular the published barycentric arrival times on 
2003 June 15, July 15 and August 15.

The Jodrell Bank ephemeris for 2003 July 15 was initially used. This gives a 
rotation frequency of 29.8044286490~Hz, a frequency derivative 
of $-3.7353779 \times 10^{-10}$
and a DM of 56.76~cm$^{-3}$pc. The standard timing package, TEMPO,
was used with this ephemeris and the arrival times for RXTE, Jodrell Bank
and Parkes data as described above. The resultant residual is shown
in the top panel of Fig~\ref{crab_fit}.
We then fitted for frequency, frequency derivative and
the frequency second derivative and the residual from this fit is shown
in the lower panel of the figure.
It can be seen that all 3 data sets agree within the errors and the radio
giants have a relatively large jitter in their arrival times.
We therefore believe that the times-of-arrival
of the RXTE data are correct within the quoted errors and that the
Parkes data are correctly de-dispersed and time tagged.
\begin{figure}
\centerline{\psfig{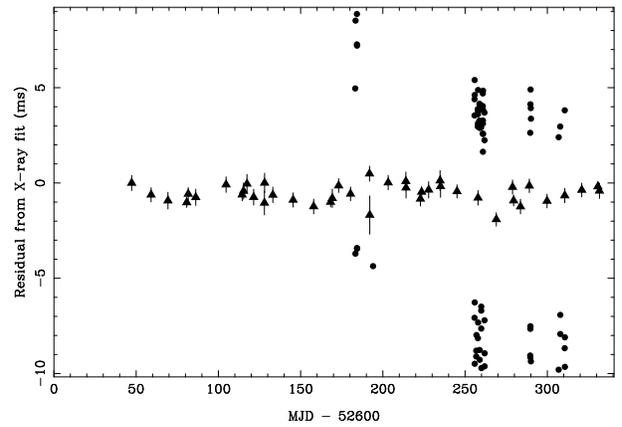}}
\caption{Timing residuals from the X-ray TOAs and giant pulse TOAs
from \psr{} compared to the model given in the third column
of Table~\ref{timtab}. Triangles denote X-ray data and circles
the radio data. The division into two clumps of giants can clearly be seen.}
\label{timing}
\end{figure}

\subsection{\psr{}}
Each dataset on \psr{} was searched for giant pulses and their
topocentic time of arrival at 1390~MHz was recorded
with an accuracy of 80~$\mu$s.
As shown in Paper~I, the giant pulses in \psr{} occur in two
groups separated by about 0.25 phase (12~ms). Within each group the
jitter is 0.1 phase (5~ms).

Simultaneous RXTE and Parkes observations were made on 2003 August 6.
The arrival of the X-ray pulse at the barycentre occurred 74832.4986 seconds 
after MJD 52857.0 and at this epoch the rotation frequency was 19.775530~Hz.
A second RXTE epoch was obtained on 2003 August 17.
Using the rotation frequency
of 19.775350~Hz obtained on that date, we computed a frequency derivative of
$-1.87\times 10^{-10}$~s$^{-2}$ between the two dates.
This yielded an initial timing solution for the pulsar. We then
included the times of arrival of the giant pulses detected during 
the August session.
For timing purposes, we picked only the largest
amplitude pulses so that the results would not be affected by possible
spurious giants at low signal to noise ratios. We split the giants into the two
groupings and timed each group separately.  This allowed us to update
the timing solution. We then added all the data obtained during
the 130 days of observing the pulsar. At this point it was necessary
to fit for the second frequency derivative.
Finally, we also obtained the TOAs for all RXTE observations of the pulsar
during 2003. We fitted the X-ray only data for the rotation frequency and
the first two derivatives.

The timing solutions are given in Table~\ref{timtab}. The pulsar position
(obtained from optical measurements by Caraveo et al. 1992\nocite{cbmm92})
and the DM (from Paper~I) were held fixed in the fitting process.
The errors are quoted
in the last digit and are twice the formal errors given by the TEMPO
software. The two groups of giants give the 
same results within the errors.
The fitted parameters obtained from the X-ray data are largely consistent
with the radio data. The frequency second derivative is the main
difference, but this term is dominated by short-term timing noise in
the radio which is averaged out in the longer X-ray data set.
As part of the fitting process we can obtain the offset between the
fiducial point of the X-ray profile and the mean location of the giants.
The late giants arrive 5.0~ms after the X-rays and the early giants arrive
6.7~ms before the X-rays. The braking index is obtained from the spin
frequency and the first two derivatives and is 
2.36 for the X-ray data consistent with the previously known result.

Fig~\ref{timing} shows the residuals for the X-ray TOAs and
the radio giant pulse TOAs compared to the model given by a fit to
the X-ray data over a period of 300 days.  In this plot,
the triangles denote the X-ray timing points. Points at positive residuals
denote the late arriving giants; those with negative residuals
denote the early arriving giants.

\section{Statistics of the giant pulses}
\begin{figure}
\centerline{\psfig{figure=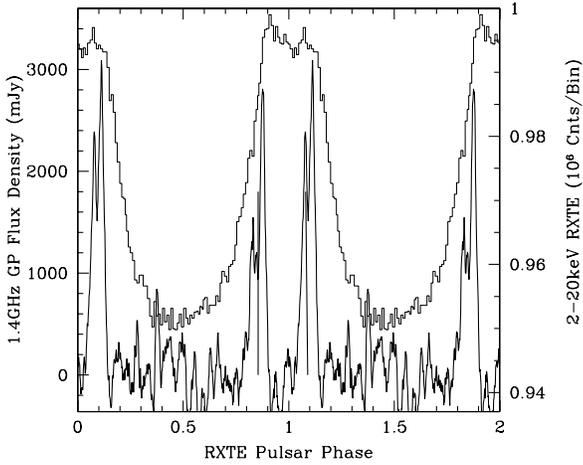,angle=0,width=8cm}}
\caption{The sum of brightest 18 giant pulses from the 2003 August
observations of \psr{} (thick line), compared with
the RXTE light curve (thin line) taken from de Plaa et al. (2003). 
Note that phase 1.0 is in the centre of the plot.
}
\label{giantprof}
\end{figure}

To highlight the relative
phases of the X-ray and radio giants, we summed together the 18 largest giant
pulses which occurred during the 2003 August observations.
We define phase zero as the peak of the sinusoidal 
X-ray timing template which in
turn is related to the high-statistics,
integrated RXTE profile of de~Plaa et al. (2003) by a fit to
the analytic decomposition of this profile. Together with
the absolute timing between the radio and X-rays as described in
Section 3, this analysis determines the relative phase 
to better than 1~ms ($\delta \phi = 0.02$).
The two profiles are shown in Fig~\ref{giantprof} where it can be
seen that the profiles are well aligned but
the peaks of giant pulse profile slightly lead the 
centroids of the corresponding X-ray pulse components.
In fact, the sharp features on the X-ray peaks
noted by de~Plaa et al. (2003; at
$\phi$= 0.94 and 0.04 in our figure)\nocite{pkh03} correspond somewhat better
to the giant profile separation and phase. 

\begin{figure}
\centerline{\psfig{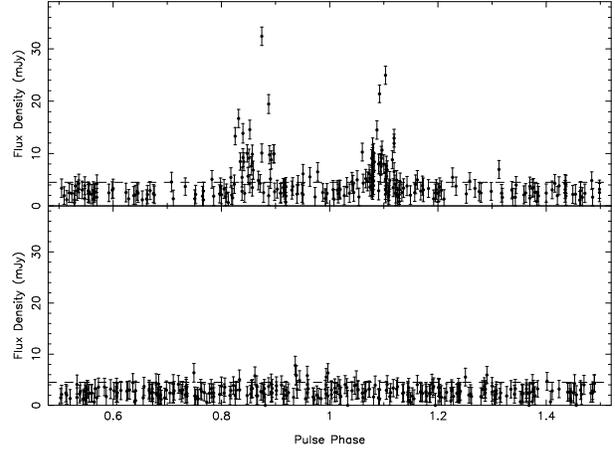}}
\caption{Location and amplitudes of giant pulses candidates from data 
taken in 2003 August.  Top panel shows data at the DM of the pulsar, the
lower panel shows the same for a DM of 125 cm$^{-3}$pc.
Note that phase 1.0 is in the centre of the plot.}
\label{location}
\end{figure}

For the data taken in 2003 August, we de-dispersed the data at the
DM of the pulsar, 146.5~cm$^{-3}$pc, and also at a DM of 125~cm$^{-3}$pc.
We used the latter data as a check on the detection algorithm for the
giant pulses as we should not have any giants at this DM.
From this, we established a detection threshold of 4.5~mJy for the giant
pulses. Figure~\ref{location} shows the flux density of giant
pulse candidates as a function of pulse phase for the data de-dispersed at both
146.5 and 125~cm$^{-3}$pc. At a DM of 125~cm$^{-3}$pc only 7 pulses
were above the threshold (out of 2 million pulses observed). At the
correct DM of 146.5~cm$^{-3}$pc we see a concentration of giant pulses
at two distinct phases.

\begin{figure}
\centerline{\psfig{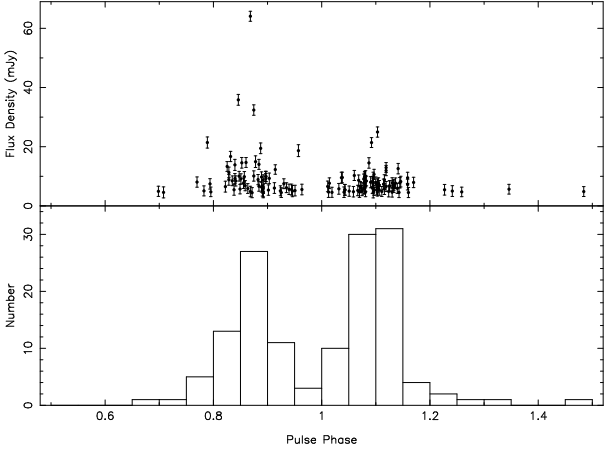}}
\caption{Top panel: Location and flux density of the 141 giant pulse
candidates obtained during 80~h observing. Bottom panel: Histogram of the 
locations of the giant pulse candidates showing their 
division into two groups. Note that phase
1.0 is in the centre of the plot.}
\label{loc2}
\end{figure}

Fig~\ref{loc2} shows the flux density versus phase of all 141 giant
pulses from the entire data set (top panel) and a histogram of location
of the giant pulses as a function of pulse phase (bottom panel).
Giant pulses seem to originate throughout the window from phases 0.78 to 1.18
although they are concentrated in two distinct peaks.
There are 60 giant pulse candidates
which arrive early (with respect to the X-ray
profile midpoint) and 81 which arrive late. If we take the rate between
phases 0.18 and 0.78 in Fig~\ref{loc2} as `background' we estimate that
no more than 3 pulses are false positives in each of the early and
late phase windows.  The early arriving giants
have a mean flux density of 10.4~mJy, higher than the 7.5~mJy
for the mean flux density of the late arrivals. The three strongest
giants from the sample are part of the early group.

\begin{figure}
\centerline{\psfig{figure=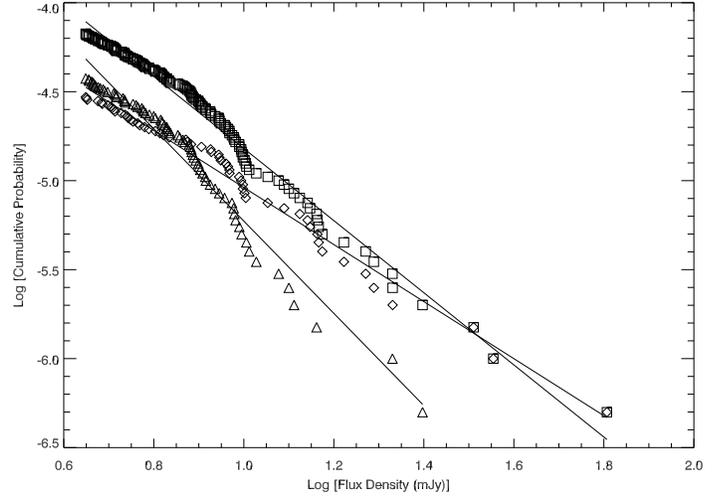,angle=0,width=10cm}}
\caption{Cumulative Log N - Log S for the giant pulses.
The `early' giants are denoted by the diamonds, the `late' giants
by the triangles and the combined data set by the squares.
A slope of $-1.8$ fits the full distribution.
The `early' giant pulses however, show a shallower slope
than the `late' giant pulses and dominate the largest events.}
\label{logns}
\end{figure}

Fig~\ref{logns} shows the unbinned cumulative probability distribution of our giant
pulses on a log-log plot. A Kolmogorov-Smirnov comparison of the
early and late giants gives a 1\% probability that these are drawn from
the same distribution. Accordingly we have used a Bayesian 
approach \cite{whe04} to fit
a power law distribution to the early and late giants separately, as
well as to the combined set of giant pulses.
These best-fit power laws are plotted on Figure~\ref{logns}.
To facilitate comparison with other pulsars we express these
giant pulse intensity distributions in units of $20\times$ the 
average pulse flux
$S_{20\times} = S_{GP}/(20\langle S_{{\rm 1.4GHz}} \rangle )$,
since this is often taken as a threshold defining giant pulse emission.
For \psr{} $20\langle S_{{\rm 1.4GHz}} \rangle= 500$~$\mu$Jy.
With these units, the best-fit
cumulative probability intensity distributions are
\begin{equation}
\begin{array}{ll}
{\rm [Early]} f_> = & 2.7 \times 10^{-4} S_{20\times}^{-1.5\pm 0.2} \\
{\rm [Late~]} f_> = & 1.4 \times 10^{-3} S_{20\times}^{-2.1\pm 0.3} \\
{\rm [Total]} f_> = & 1.2 \times 10^{-3} S_{20\times}^{-1.8\pm 0.2}
\end{array}
\end{equation}
The limited statistics and refractive variations imply a $\sim20$ per cent
uncertainty in the normalizations. For comparison, the Crab giant pulses have
$f_> = 6 \times 10^{-3} S_{20\times}^{-2.3}$ (at 0.8~GHz; Lundgren et al 1995)
while PSR~B1937+21 shows giant pulses with
$f_> = 1 \times 10^{-5} S_{20\times}^{-1.8}$
(at 1.4~GHz, Kinkhabwala \& Thorsett 2000\nocite{kt00}).
Note that the giant pulse distributions in \psr{}
show some evidence for a break to a flatter
index for $S < 7~\mu$Jy, but with a limited dynamic range it is
difficulty to quantify the significance of such a break. There is no
evidence of a cut-off at large pulse amplitudes. As noted below there is
some evidence for refractive interstellar scintillation modulation of
both the integrated profile flux and the giant pulse rate. We do not
have sufficient signal-to-noise to calibrate this modulation. It does
not strongly affect the fitted pulse distributions, although we note that
largest two (early) giant pulses were seen at periods of low integrated flux.
Correction for this effect would tend to further flatten the early giant pulse
distribution.

The largest pulse (from the early group) has an equivalent
continuum flux of 64~mJy, some 2500$\times$ the mean flux and 3800$\times$
the integrated flux density for this particular observing epoch.
Overall, we obtain a $\gapp$10~mJy pulse every two hours, about equally from
the two components. We expect one $\gapp$100~mJy pulse 
in $\sim$9~d of observing; this will likely occur from the early component.

\section{Integrated profile of \psr{}}
\begin{figure}
\centerline{\psfig{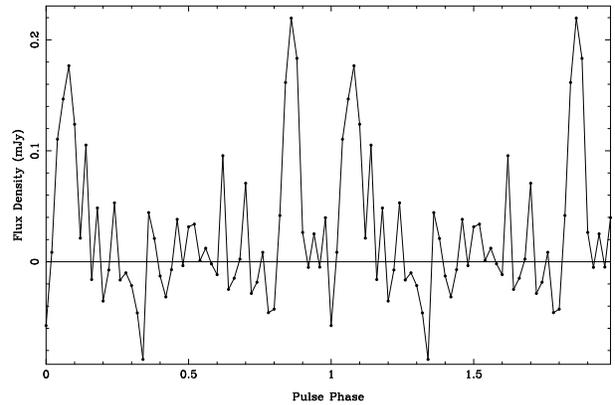}}
\caption{Integrated profile of \psr{} at 1.4~GHz from observations 
taken in 2003 August. Total time on-source was 31.2 h.
There are 50 bins across the
pulse phase for a resolution of $\sim$1 ms. The rms is 33~$\mu$Jy.}
\label{integprof}
\end{figure}
Armed with a timing solution we can now sum the data together to determine
the integrated profile of \psr{}. Fig~\ref{integprof} shows
the profile after summation of the data taken in 2003 August.
The pulsar is clearly detected with an integrated flux density
of 24~$\mu$Jy. We also formed the integrated profile for each of
the other observing sessions from 2003 and from 2001.
With less integration time, the
signal to noise ratio is not as good as in Fig~\ref{integprof} but the flux
densities we obtain are 16~$\mu$Jy (2001), 19~$\mu$Jy (May/June 2003)
and 20~$\mu$Jy (Sept 2003). We note that the
variability in these values is consistent
with the giant pulse rate which also varies slightly from session to session.
There is therefore some evidence
for variability in the flux density with a modulation index of 0.15,
likely due to refractive
scintillation effects where the expected timescale is of order days.
Given the 640~MHz flux density quoted by
Manchester et al. (1993)\nocite{mml+93},
the radio spectral index of the integrated emission between 640 and
1390~MHz is $-3.6$, one of the steepest of all the known pulsars.
In the Crab pulsar, the main pulse has a spectral index of $-3.0$,
however this spectral index is a mixture of the `normal' radio emission
which likely has a steeper spectrum again and the flatter spectrum
giant pulse emission \cite{sbh+99}.

The integrated profile looks similar to the `giant profile' shown
in Fig~\ref{giantprof}, with a peak ratio similar to that seen in our
set of early and late giants. However, the individual giants that
we detect above threshold contribute only 0.2~$\mu$Jy per pulse or 
less than 1\% of the integrated flux density. One may ask whether
a population of sub-threshold giants makes up most of the integrated
profile (as for the Crab main and interpulse above 800~MHz;
Lundgren et al. 1995\nocite{lcu+95})
or whether the integrated profile is dominated by a `normal' pulse
population with a $\sim$ Gaussian intensity distribution, as for
PSR~B1937+21, where the giant pulse emission is less than 1\% of the total
\cite{kt00,jap01}.

A cumulative giant pulse number distribution $f_> = A s^\alpha$ contributes
a flux $\langle s \rangle = \alpha\,\,A\,\,s_0^{(\alpha+1)}/(\alpha+1)$ above
some cut-off $s_0$. For our intensity distributions above, even if every
early pulse was drawn from the giant distribution (down to a threshold
of $2~\mu$Jy), the giant pulses would only account for half of the
first component flux. The late component could, in principal, be made up
of giant pulses representing 10 per cent of the pulses, with a flux threshold
of $\sim 65~\mu$Jy $\approx 2.5 \langle s \rangle$. Given the similarity
in the number of early and late giants above our observation threshold
this seems rather unlikely. Instead, it seems more likely that only a modest
fraction of the pulses are giants as for the Crab and PSR~B1937+214. If
we adopt a giant pulse fraction of $10^{-3}$, then the flux threshold for a giant pulse
in the early component is $200~\mu$Jy and these contribute 0.6~$\mu$Jy
or 5 per cent of the early component flux. The equivalent numbers for the late
component give a threshold of $\sim 600~\mu$Jy and an integrated flux of
$1.1~\mu$Jy, or $\sim 10$ per cent of the late component flux. This would imply
that both components are dominated by a conventional pulse distribution.
Given the low flux, this will be difficult to confirm.

\section{Radio and X-ray comparison}
Although we have made several attempts to coordinate Parkes and
RXTE observations, and have obtained several near-simultaneous epochs,
only a modest amount of integration was simultaneous during the 2003
campaign. However, three giant radio pulses were obtained during an
RXTE exposure on 2003 August 6 ($\sim$MJD 52857.9) and two were observed on
2003 September 28 ($\sim$MJD 52910.6). Examination of the
X-ray data stream during these
large ($\ga 10$~mJy) pulses allows an initial search for a high energy excess.
We have co-added the X-ray events surrounding the infinite
frequency barycentric arrival time of these pulses. We have examined the
RXTE light curve on a range of timescales, finding no excess counts associated
with the radio giant pulses.

We base our fiducial limit on the 5$\times$5~ms time intervals centered on the 
giant pulses.  The number of events expected from background, 
including diffuse X-rays, unrejected cosmic rays, X-rays from 
nearby sources which are in the collimated field of view, as
well as regular pulses from the pulsar, is estimated to be 2.1
in these 25~ms.  Since only 4 events are observed
during these five windows, there is no statistically 
significant excess that can be associated with the giant pulses.
Using Poisson statistics, we calculate that at the 90 per cent
(99.5 per cent) confidence level
the actual excess in coincidence with giant radio pulses is no 
more than 6.0 (10.6) events. In comparison, the phase-averaged X-ray 
pulse rate produces 0.084 counts in the interval. In other words,
at the 90 per cent confidence level, any giant X-ray pulses associated 
with the radio giant pulses are no larger than $71\times$ the normal pulses.

However, considering that these detected radio giants represent
an average excess of $400\times$ the mean radio emission, it is interesting
to note that the X-ray enhancement is constrained to be $5.5\times$ smaller
than that in the radio. These results are consistent with other
comparisons of high energy fluctations with radio giant pulse
emission as detailed in the introduction. It is evident that the factors of
1000 or more seen in the differences between the giant pulse emission
and the normal emission in the radio are not observed at higher energies.

The limits on the high energy pulse modulation and
the rather small claimed enhancement on the optical flux associated with
radio giant pulse events
certainly suggest that giant pulses are caused by
coherence fluctuations rather than changes in total plasma density.
Accordingly it is intruiging that the giant pulse phenomenon appears to be associated
with the light cyclinder field rather than the surface dipole. This might
suggest that growing instabilities (e.g. two-stream or radiation back-reaction)
in plasma flowing through the outer magnetosphere mediate the radio giant pulse
modulations, and that the longer path-length in this zone allows
larger fluctuation amplitudes.

\psr{} also apparently follows the Crab and the millisecond
pulsars PSR~B1937+21 and B1821--24 in having the giant pulse
emission confined to phases that also
show narrow, hard-spectrum X-ray components. Better statistics and
improved relative timing might show that the giant pulse events are associated with
the peak structure in the X-ray pulse of \psr{}, which would further
strengthen this conclusion.

If we take this connection with the outer magnetosphere and with
the hard X-ray emission seriously, we should ask how this relates to the
expected pulse structure of high energy emission models. Both the outer
gap picture described by Romani \& Yadigaroglu (1995)\nocite{ry95}
and the two-pole
(extended slot gap) picture of Dyks, Harding \& Rudak (2004)\nocite{dhr04} give
loci for narrow pulse components far from the star surface that lie
off the magnetic pole. At present, the best clue we have to the emission
geometry in \psr{} is the 0.25 phase separation of the giant pulse components.
Such separations are naturaly seen in the outer gap picture for
viewing angle $\zeta \le 75^\circ$. This may, however,
be in disagreement with the
X-ray torus of wind nebula around \psr{} as measured by the Chandra X-ray
Observatory \cite{gw00}, which
appears to be closer to edge-on $\zeta \approx 85^\circ$. It is not clear
if the two-pole models can give such narrow pulse separations for any
angles. Certainly independant measurement of the viewing geometry would
help greatly in constraining the location of the giant pulse emission regions.
Unfortunately, radio polarization measurements to test the viewing angles
will be difficult to obtain for this pulsar; the position 
angle sweep may however be measurable in the optical band.

\section{Conclusions}
We have carried out extensive observations of \psr{} at radio wavelengths and
have also obtained simultaneous radio and X-ray observations of the pulsar.
The observations have enabled us to determine that the giant pulses in the 
radio arrive in-phase with the X-ray pulses. However, we see no enhancement in
the X-ray flux at the time of the radio giants.
We have shown that the giant pulses follow power-law statistics. It seems likely
that they contribute only a few percent to the integrated flux density.
All four pulsars which are giant pulse emitters (two young pulsars and two 
millisecond pulsars) show similar characteristics.

\section*{Acknowledgments}
RWR is supported in part by NASA grant NAS-13344.
The Australia Telescope is funded by the Commonwealth of 
Australia for operation as a National Facility managed by the CSIRO.
We thank J.~Reynolds for providing unallocated telescope time 
for this project and A.~Karastergiou and S.~Ord for help with observing.

\bibliography{modrefs,psrrefs,crossrefs}
\bibliographystyle{mn}
\label{lastpage}
\end{document}